\PassOptionsToPackage{draft}{hypref}
\PassOptionsToPackage{breaklinks=true}{hypref}
\documentclass[a4paper,fleqn,usenatbib]{mnras}

\usepackage{graphicx}
\usepackage{grffile}
\usepackage{natbib}
\usepackage{hyperref}

\newcommand{\RNum}[1]{\uppercase\expandafter{\romannumeral #1\relax}}

\newcommand{\rec}{\ensuremath{\bar{r}_{e}}}

\newcommand{\uae}{\ensuremath{\bar{\mu}_{e}}}
\newcommand{\axrat}{\ensuremath{q}}
\newcommand{\prerec}{\texttt{R50}}
\newcommand{\preuae}{\texttt{SB\_N50}}

\newcommand{\SELrecL}{\ensuremath{3\arcsec}}
\newcommand{\SELrecU}{\ensuremath{10\arcsec}}
\newcommand{\SELuaeL}{\ensuremath{24}}
\newcommand{\SELuaeU}{\ensuremath{27}}
\newcommand{\SELnU}{2}
\newcommand{\SELuniquedist}{5\arcsec}
\newcommand{\prerecL}{\ensuremath{1\arcsec}}
\newcommand{\preuaeL}{\ensuremath{23}}

\newcommand{\MLnfeatures}{26}
\newcommand{\MLerrtol}{20\%}

\newcommand{\SRCnpresel}{14108}
\newcommand{\SRCnselnohuman}{652}
\newcommand{\SRCnsel}{479}
\newcommand{\SRCnselr}{225}
\newcommand{\SRCnselb}{254}
\newcommand{\SRCnselgama}{215}
\newcommand{\SRCnselgamar}{88}
\newcommand{\SRCnselgamab}{127}
\newcommand{\SRCnselrperc}{47}
\newcommand{\SRCnselbperc}{53}
\newcommand{\SRCnselgamarperc}{41}
\newcommand{\SRCnselgamabperc}{59}
	
\newcommand{\MTObjects}{{\scshape MTObjects}}

\newcommand{\SExtractor}{{\scshape SExtractor}}
\newcommand{\Imfit}{{\scshape imfit}}
\newcommand{\DeepScan}{{\scshape DeepScan}}
\newcommand{\ProFound}{{\scshape ProFound}}
\newcommand{\Eazy}{{\scshape EAZY}}

\newcommand{\patchmeandepth}{26.2}
\newcommand{\patchnumber}{12767}
\newcommand{\patcharea}{480}
\newcommand{\patchrejpercent}{30}

\newcommand{\SynthRangeRec}{1\arcsec$<$\rec$<$15\arcsec}
\newcommand{\SynthRangeUae}{22.0$<$\uae$<$28.5}
\newcommand{\SynthRangeN}{0.2$<$$n$$<$2.5}
\newcommand{\SynthRangeQ}{0.1$<$\axrat$\leq$1.0}
\newcommand{\SynthNumber}{$\sim$300,000}
\newcommand{\REparam}{\ensuremath{\epsilon_{r}(\rec, \uae)}}

\newcommand{\FitSizeMinPix}{200}
\newcommand{\FitSizeMaxPix}{500}
\newcommand{\FitSizeMinArcsec}{34}
\newcommand{\FitSizeMaxArcsec}{85}

\newcommand{\GAMAstellarmasses}{\texttt{StellarMassesv20}}
\newcommand{\GAMAcolourcat}{\texttt{ApMatchedCatv06}}

\newcommand{\zphotmin}{$z$=$0.001$}
\newcommand{\zphotmax}{$z$=$1$}

\newcommand{\dk}{\ensuremath{\bar{r}_{k}}}

\newcommand{\RESqf}{\ensuremath{26\pm5\%}}
\newcommand{\RESqfupper}{\ensuremath{30\%}}

\author[D. J. Prole et al.]{D. J. Prole,$^{1, 2}$\thanks{daniel.prole@mq.edu.au}
	R. F. J. van der Burg,$^{3}$
	M. Hilker$^{3}$
	and L. R. Spitler$^{1, 2}$
	\\
	$^{1}$Research Centre for Astronomy, Astrophysics \& Astrophotonics, Macquarie University,
	Sydney, NSW 2109, Australia\\
	$^{2}$Department of Physics \& Astronomy, Macquarie University, Sydney, NSW 2109, Australia\\
	$^{3}$European Southern Observatory, Karl-Schwarzschild-Str. 2, 85748 Garching bei M\"unchen, Germany \\
}

\title[Low surface brightness galaxies]{The quiescent fraction of isolated low surface brightness galaxies: Observational constraints}

\date{}

\pubyear{2018}

\begin{document}
	\label{firstpage}
	\pagerange{\pageref{firstpage}--\pageref{lastpage}}
	\maketitle
	

\begin{abstract}

Understanding the formation and evolution of low surface brightness galaxies (LSBGs) is critical for explaining their wide-ranging properties. However, studies of LSBGs in deep photometric surveys are often hindered by a lack of distance estimates. In this work, we present a new catalogue of \SRCnsel\ LSBGs, identified in deep optical imaging data from the Hyper Suprime-Cam Subaru Strategic Program (HSC-SSP). These galaxies are found across a range of environments, from the field to groups. Many are likely to be ultra-diffuse galaxies (UDGs). We see clear evidence for a bimodal population in colour - S\'ersic index space, and split our sample into red and blue LSBG populations. We estimate environmental densities for a subsample of \SRCnselgama\ sources by statistically associating them with nearby spectroscopic galaxies from the overlapping GAMA spectroscopic survey. We find that the blue LSBGs are statistically consistent with being spatially randomised with respect to local spectroscopic galaxies, implying they exist predominantly in low-density environments. However, the red LSBG population is significantly spatially correlated with local structure. We find that \RESqf\ of isolated, local LSBGs belong to the red population, which we interpret as quiescent. This indicates that high environmental density plays a dominant, but not exclusive, role in producing quiescent LSBGs. Our  analysis method may prove to be very useful given the large samples of LSB galaxies without distance information expected from e.g. the Vera C. Rubin observatory (aka LSST), especially in combination with upcoming comprehensive wide field spectroscopic surveys.

\end{abstract}



\section{Introduction}

Low surface brightness (LSB) galaxies have $r$-band surface brightnesses fainter than around 24 magnitudes per square arc-second averaged within their effective radii, with stellar masses less than 10$^{9}$M$_{\odot}$ \citep{Wright2017}. However, they are larger and more diffuse than dwarf galaxies in the same mass range. Despite LSB galaxies (LSBGs) being among the most numerous in the Universe \citep{Martin2019}, they also remain among the most poorly understood. This is because of the relative difficulty in detecting and measuring such faint objects in modern astronomical surveys that arises not only from their diffuse nature, but also from systematic limitations in the instrumentation \citep[e.g.][]{Abraham2014}, observing technique \citep[e.g.][]{Trujillo2016}, data reduction approach \citep{Fliri2016} and source extraction software \citep{Akhlaghi2015, Prole2018} in the LSB regime. 

\indent LSB galaxies have been studied for decades \citep[e.g.][]{Sandage1984, Bothun1991, Sabatini2003}, but the majority\footnote{Although exceptions exist, e.g. \cite{McGaugh1995, Dalcanton1997}.} of this work has focused on those in galaxy groups and clusters where their distances can be easily estimated by associating them with the dense environment. Estimating distances for galaxies outside galaxy groups is more difficult; long exposure times are often required to obtain precise redshift estimates, meaning that spectroscopic galaxy surveys typically suffer drastic incompleteness in the LSB regime \citep[e.g.][]{Wright2017}. Statistical constraints on the properties of these galaxies are therefore sparse among the literature. This is particularly true for more isolated LSB galaxies whereby distance estimation based on association with nearby galaxy groups is not possible. This selection bias restricts scientific understanding of the formation and evolutionary pathways of such objects because disentangling the role of environment is impossible without knowing their properties outside of dense environments.

\indent LSB galaxies are typically dwarf galaxies in terms of their stellar masses ($M_{\star}$$\leq$$10^{9}$M$_{\odot}$) and metallicities, but are known to span a wide range in halo mass \citep{Beasley2016, Mowla2017, vanDokkum2018, Prole2019a} and physical size; there are now many known examples of very large ($r_{e}$$>$1.5 kpc) LSB galaxies \citep[i.e. ultra-diffuse galaxies or UDGs][]{vanDokkum2015}. This variety in the properties of LSB galaxies is suggestive of an equally diverse range in their formation and evolutionary channels, which can be secular - driven by processes internal to the galaxy, or caused by interactions with their external environment.

\indent Environmentally-driven formation usually refers to interactions with other galaxies, the group or cluster potential, or the gas between galaxies in dense environments. The environment plays an important role in galaxy evolution, turning disky, irregular dwarf galaxies into pressure-supported spheroidal systems \citep{Kazantzidis2011}. LSB galaxies can also form by quenching star formation in a galaxy early on, for example during cluster in-fall \citep{Yozin2015}. Aside from their range in sizes, this can also explain why they are generally quiescent inside galaxy clusters \citep[e.g.][]{Koda2015, vanderBurg2017}, following the general trend of decreasing early-type fraction with halo-centric radius observed for the general galaxy population \citep{Weinmann2006}. Moreover, \cite{Carleton2019} have demonstrated that tidal stripping and heating of dwarf galaxies with cored dark matter profiles can produce LSB galaxies and is sufficient to reproduce the sizes and stellar masses of UDGs in clusters \citep[also see e.g.][]{Collins2013}. Relative depletions in the numbers of dwarf galaxies (including UDGs) towards the centres of galaxy clusters have been observed \citep[e.g.][]{vanderBurg2016, Venhola2017, Pina2018}, suggesting that extremely dense environments can be destructive for such objects \citep[e.g.][]{Janssens2019}. It is also possible for LSB galaxies to form from the tidal debris left over from galaxy mergers \citep{Ogiya2018, Bennet2018}, although the resulting systems ought to be higher in metallicity than typical LSB galaxies and may be dark matter deficient. 

\indent Alternatively, secular evolutionary channels can impact even isolated galaxies. Secular processes include supernovae feedback, which can cause expansion of the stellar haloes (and thereby dimming of surface brightness) in dwarf galaxies via outflows \citep{DiCintio2017, Chan2018}, and the development of extended stellar components from higher than average angular momentum in the halo \citep{Jimenez1998, Amorisco2016, Rong2017}, supported by rotation measurements of H\RNum{1}-rich UDGs \citep{Leisman2017}. These mechanisms have been shown to reproduce several observed properties of LSB galaxies such as their sizes, stellar masses and morphologies. 

\indent The relative significance of the various formation mechanisms is not well understood for the global population of LSB galaxies. For instance, \cite{Martin2019} have shown that dwarf galaxies can undergo rapid star formation in the early Universe, which results in shallow stellar profiles which are more easily influenced by tidal forces. \cite{Jiang2019} find that theoretically half of the UDGs in groups could be created from tidal heating of dwarf galaxies, whereas the other half originate from outside the group environment, becoming redder during in-fall because of ram pressure stripping. This process can account for the observation that UDGs may be systematically bluer \citep{Roman2017} towards the outskirts of groups. 

\cite{Prole2019b} have shown that UDGs are indeed systematically bluer in the field and perhaps occupy a similar mass fraction as those in groups and clusters. However, a small number of quiescent UDGs have been found to exist in isolation \citep{Delgado2016, Roman2019}. The origins and prevalence of this type of galaxy are unknown, thanks in large to the difficulties of detecting red LSB objects, which are generally fainter than their bluer counterparts \citep[e.g.][]{Greco2018, Tanoglidis2020} and are much harder to distinguish from background galaxies. As such, there has been a relative absence of discussion surrounding these objects in the recent literature, from either an observational or theoretical perspective. 

\indent Understanding these objects is crucial for understanding the secular processes that lead to quenched star formation in isolated dwarf galaxies, which may shed light on important topics such as supernovae or photoelectric feedback (i.e. that from hot young stars), the latter of which likely plays the dominant role in suppressing star formation in dwarf galaxies \citep{Forbes2016}. An interesting possibility is that some isolated quiescent LSBGs may only be in a temporary state of quiescence, owing to their bursty star formation histories \citep{Muratov2015} which can cause quenching following periodic episodes of intense star formation, which may be triggered by the accretion of either new gas or that which was previously expelled. 

\indent The goal of this study is to constrain the quiescent fraction of isolated LSBGs, in order to understand this often-overlooked population. Surface brightnesses are always given in units of magnitudes per square arc-second and we use the AB magnitude system. Cosmological calculations are performed assuming $\Lambda$CDM cosmology with $\Omega_{\mathrm{m}}$=0.3, $\Omega_{\Lambda}$=0.7, H$_{0}$=70 kms$^{-1}$Mpc$^{-1}$. In $\S$\ref{section:data} we describe the data we use to identify LSBGs. The source catalogue is described in detail in $\S$\ref{section:observations}. We describe the method to calculate the quiescent fraction in $\S$\ref{section:analysis}. We discuss and conclude in sections \ref{section:discussion} and \ref{section:conclusions}.


\section{Data}
\label{section:data}

We use the calibrated co-added $grizy$ images provided in the second public data release (PDR2-wide) of the Hyper Suprime-Cam Subaru Strategic Program (HSC-SSP) survey data\footnote{available online: \url{https://hsc-release.mtk.nao.ac.jp/doc/}} \citep{Aihara2019} to produce a new catalogue of LSB galaxy candidates for this study. In particular, we use the $r$-band for source detection, in-keeping with similar studies \citep{vanderBurg2017, Prole2019b}. The data are already split into discrete patches approximately 12$\arcmin$$\times$12$\arcmin$ in size, with 0.168$\arcsec$ pixels. The average $r$-band seeing is 0.76$\arcsec$. The dataset has been reduced with an improved background subtraction algorithm compared to the previous release, meaning LSB structure is well preserved over $\sim$arc-minute scales. The HSC data is roughly half a magnitude deeper than the comparable Kilo-Degree Survey \citep[KiDS,][]{deJong2013} in the $r$-band.

\indent We target the W03 and W04 HSC-SSP regions in order to exploit the 180 deg$^{2}$ overlap with the three equatorial regions (G09, G12, G15) defined by the GAMA spectroscopic survey \citep{Driver2011}. The depth of the data is not uniform over this area since the survey is not yet complete. In order to obtain a more homogeneous dataset, we reject the bottom \patchrejpercent\% of HSC patches in terms of their 5$\sigma$ point source depth in the $r$-band, as measured by the HSC-SSP team. We process \patchnumber\ patches ($\sim$\patcharea\ deg$^{2}$) in total, the mean $r$-band depth of which is \patchmeandepth\ mag.
 

\section{Observations}
\label{section:observations}

\subsection{Source Extraction Pipeline}

\indent We do not use the official HSC-SSP source catalogues to identify LSB sources because of the need for specialised source extraction techniques in the LSB regime \citep[e.g.][]{Prole2018} and because it is necessary for us to quantify our ability to recover such sources in order to correct for incompleteness effects. Our adopted source extraction pipeline consists of the following steps, which are expanded upon in subsequent sections:

\begin{enumerate}
	\item First, we run the \MTObjects\ \citep[][]{Teeninga2015} over the data to produce an initial source catalogue and segmentation image for each HSC-SSP patch. A preselection is applied to these sources based on their segment properties before each is measured in detail. For details, see $\S$\ref{section:sourceextraction}.
	
	\item We fit 2D S\'ersic models to each preselected source using the \Imfit\ \citep{Erwin2015} software.  LSBG candidates are selected using the resulting structural parameters combined with other measurements. This is described in detail in $\S$\ref{section:structural}.
	
	\item Colours are measured for the selected sources using all five bands and elliptical annuli, as described in $\S$\ref{section:colours}.
\end{enumerate}

\indent We define our morphological selection criteria as \SELrecL$<$\rec$<$\SELrecU, \SELuaeL$<$\uae$<$\SELuaeU\ and $n$$<$\SELnU, where \rec\ is the circularised effective radius\footnote{While there is some controversy over using the effective radius to compare the sizes of UDG-like galaxies to e.g. the Milky Way \citep[e.g.][]{Chamba2020}, we can still make fair comparisons between galaxies of similar morphologies as are anticipated for LSB galaxies.}, \uae\ is the mean surface brightness within the effective radius, and $n$ is the S\'ersic index \citep[e.g.][]{Graham2005}, a proxy for morphology. This latter limit is justified because LSB galaxies including UDGs are typically observed with $n$=1 \citep[e.g.][]{Davies1988, Koda2015}, as expected theoretically \citep[e.g.][]{DiCintio2017}. These selection criteria are justified and  expanded-upon in $\S$\ref{section:srccat}. The source extraction pipeline is discussed in further detail throughout the remainder of this section.
	

\subsubsection{Source Extraction}
\label{section:sourceextraction}

\indent We use \MTObjects\ for source identification because of its improved performance in the LSB regime compared to other software such as \SExtractor\ \citep{Bertin1996, Prole2019b}. Its main outputs are a segmentation image and a catalogue of sources with measurements based on the pixels contained within each segment. 

\indent However, we found that the source measurements produced by \MTObjects\ alone were severely biased in the LSB regime based on artificial source injections. We choose to supplement our source catalogue with the following additional segment statistics based on their equivalents in the \ProFound\ photometry package \citep{Robotham2018}:

\begin{enumerate}
	
	\item \texttt{R50}: This is a non-parametric proxy for the circularised half-light radius. It is measured by counting the number of  pixels (sorted by flux in decreasing order) before the sum of preceding pixels exceeds the half-flux of the segment. This is equivalent to the half-light area, so can be divided by $\pi$ to obtain a proxy for \rec$^{2}$.
	
	\item \texttt{SB\_N50}: This is a proxy for the average surface brightness within the half-light radius. It is simply obtained by averaging the flux of pixels used in the calculation of \texttt{R50}.
		
\end{enumerate}

\indent While these measurements are still biased for LSB sources, we found that the effect was not so severe as for those obtained directly from \MTObjects. We preselect sources using these segment statistics in order to reduce the number of detailed (time-consuming) measurements and therefore improve the overall processing time. The preselection criteria are tuned to retain a high level of completeness in our final selection range. Specifically, we require \prerec$>$\prerecL and \preuae$>$\preuaeL. A set of artificial source tests, introduced in $\S$\ref{section:injections}, show that the vast majority of sources that are compliant with our final selection criteria also meet the preselection criteria. 


\begin{figure*}
	\includegraphics[width=\linewidth]{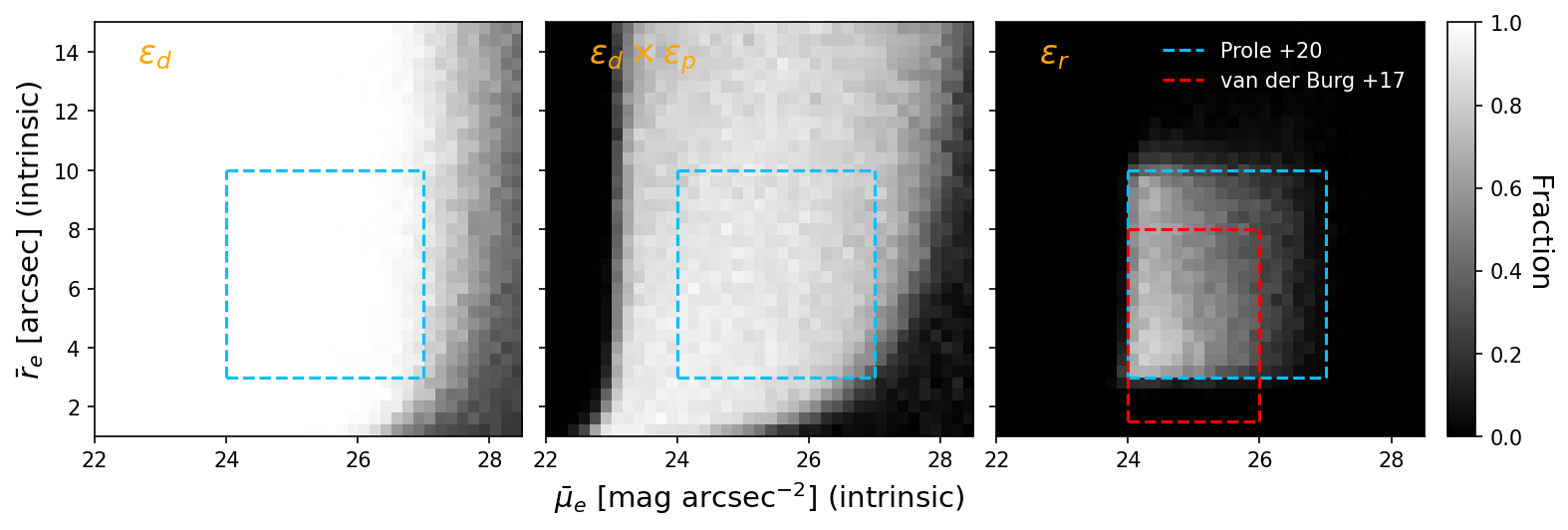}
	\centering
	\caption{The detection efficiency $\epsilon_{d}$, the combined detection and preselection efficiency $\epsilon_{d}$$\times$$\epsilon_{p}$ and overall recovery efficiency $\epsilon_{r}$ as estimated from our artificial source injections (greyscale histograms), compared to our selection criteria (blue boxes) and that of \protect\cite{vanderBurg2017} (red box), who used \SExtractor\ on the KiDS data. The colour scale shows the fractional completeness. The recovery efficiency is defined in terms of intrinsic apparent structural parameters. The blue box is the selection criteria expressed in terms of \rec\ and \uae, which is only applied in the third panel. The drop in completeness between the first and second panels is largely due to blending between the artificial sources and real ones in the images. This figure includes the effects of the machine-learned selection described in $\S$\ref{section:ML}.}
	\label{figure:RE}
\end{figure*}

\subsubsection{Structural Parameters}
\label{section:structural}

\indent Measurements based on segment statistics are often unavoidably biased for LSB objects \citep{Disney1976, Disney1983}. Parametric modelling can partially overcome these problems by taking into account additional pixels and extrapolating into the LSB regime. The usefulness of this approach depends on how well the models describe the intrinsic source morphologies. While quiescent LSB galaxies are often well-fit by simple S\'ersic models, star-forming examples can often be irregular and require additional model components. However, S\'ersic fitting is useful for the first-order size estimates required by this study.

\indent We use the \Imfit\ \citep{Erwin2015} software to fit 2D S\'ersic profiles to each preselected source. In addition to the  S\'ersic component, we simultaneously fit a flat, inclined sky model because we found it important for improving the accuracy of recovered S\'ersic parameters. Because we are targeting well-resolved sources, we neglect the PSF during the initial fitting in order to dramatically speed up the measurement process. We verify that this does not severely impact the accuracy of recovered structural parameters by means of artificial source injections $\S$\ref{section:injections}. We re-perform the fits to include the effect of the PSF for our final sample (i.e. following the application of our selection criteria; see $\S$\ref{section:srccat}).

\indent The segment measurements were used to provide initial parameter estimates for the fits, assuming \rec$\sim$\prerec, \uae$\sim$\preuae, $n$=1 and  a flat sky level of zero. Initial estimates of the axis ratios and positional angles were also based on the segment statistics. We used a minimum side of \FitSizeMinPix\ pixels (\FitSizeMinArcsec\arcsec) and a maximum of \FitSizeMaxPix\ pixels (\FitSizeMaxArcsec\arcsec) for the parametric fits. This is ample given the upper size constraint of \rec$<$\SELrecU\ in our selection criteria. If the bounding box of the detection was larger than this limit (usually because of blending with a large source nearby), we used the bounding box itself as the fitting region. Segments belonging to other sources were masked during the fitting. 

\indent There are many examples of nucleated LSB galaxies amongst the literature \citep[e.g.][]{Venhola2017}. For this analysis however, we are primarily interested in measuring the properties of the diffuse LSB stellar component. The presence of compact sources (e.g. nuclear star clusters, globular clusters) has the potential to affect the accuracy of the fits. We implemented a point source masking algorithm in order to remove pixels belonging to such objects during the fitting. The algorithm uses a high pass filter, obtained by subtracting a smoothed image  ($\sigma$=3 pixels) from the original data and masking out any bright (S/N$>$5) residuals from the result. Again, we used artificial source injections to verify that this does not bias our size measurements.


\subsection{Recovery efficiency and measurement uncertainties}
\label{section:injections}

\indent Quantifying the recovery efficiency, defined as the fraction of sources intrinsically meeting our selection criteria that survive in the final source catalogue, in addition to measurement errors (both random and systematic) is necessary to perform an analysis free from selection bias. We rely on artificial source injections to accomplish this in similar fashion to \cite{vanderBurg2016} and \cite{Prole2019b}.

\indent Ideally the artificial sources should be injected into raw data prior to any reduction such as sky subtraction. However, for this analysis we only inject them into existing reduced HSC-SSP co-adds. This does not significantly compromise the validity of the artificial source tests because the background subtraction takes place over much larger scales than the maximum size of the injected sources. Nevertheless, we apply our full source extraction pipeline, including source identification and structural parameter measurement to the augmented patches so the resulting estimate of the recovery efficiency is as realistic as possible. One notable difference is that we do not preselect sources before structural parameter measurement in order to test and optimise the preselection criteria.

\indent We inject single component S\'ersic profiles that are convolved with the PSF; we use fiducial per-patch PSF images obtained from the HSC-SSP archive. The S\'ersic parameters are randomly sampled over the following ranges: \SynthRangeRec, \SynthRangeUae, \SynthRangeN, \SynthRangeQ, where \axrat\ is the axis ratio, as well as having fully randomised position angles.

\indent We ensure that the artificial sources do not overlap with each other by positioning them in a regular grid, with each artificial source being allocated a 90\arcsec$\times$90\arcsec area. The positions of the injected sources within the grid are fixed, but we randomise the position of the grid within the image to ensure that sources can be injected at all possible positions. We do not inject sources near the borders of the images. Overall, we injected \SynthNumber\ random sources into the data to form the basis of our recovery efficiency measurement. 


\subsubsection{Recovery Efficiency}

\indent A circular area of radius 3\arcsec centred on the central position of the injected sources is used to associate them with segments in the segmentation map produced by \MTObjects. In the case of multiple matches with one injection, the segment with the largest number of pixels within the circular area is selected. At this point, any injection with a match is regarded as detected; the fraction of these sources defines the detection efficiency $\epsilon_{d}$. The preselection efficiency $\epsilon_{p}$ is similarly defined as the fraction of detected sources that meet the preselection criteria. The selection efficiency $\epsilon_{s}$ is defined as the fraction of preselected sources that meet the selection criteria. The total recovery efficiency $\epsilon_{r}$ is then calculated from the product:

\begin{equation}
\epsilon_{r} = \epsilon_{d}\times\epsilon_{p}\times\epsilon_{s}
\label{equation:RE}
\end{equation}

\indent The injections can be used to parametrise the recovery efficiency as a function of intrinsic (i.e. free from measurement error) apparent S\'ersic parameters, \REparam. Note that we neglect $n$ and \axrat\ from this parametrisation for simplicity as $n$ is expected to be distributed over a narrow range and \axrat\ is already partially accounted for by using the circularised effective radius. Equation \ref{equation:RE} is visualised using this parametrisation in figure \ref{figure:RE}. 

\indent We use the artificial source injections combined with the 5$\sigma$ point source patch depths to quantify how the recovery efficiency varies between frames. On average, the overall recovery efficiency is found to systematically vary by $\pm$2\% from the results shown in figure \ref{figure:RE}. This variation is not significant enough to impact our analysis and we adopt a fiducial recovery efficiency for the whole footprint based on this.


\subsubsection{Measurement uncertainties and Machine-learned selection}
\label{section:ML}

\indent The recovered structural parameters of the injections are used to quantify the measurement biases and uncertainties. The advantage of this approach is that any systematic errors can be directly accounted for in addition to the random errors. Since the injections are spread uniformly over the full dataset, these uncertainties also statistically incorporate the effects of blended sources and field crowding on the measurement accuracy and precision.

\begin{figure}
	\includegraphics[width=\linewidth]{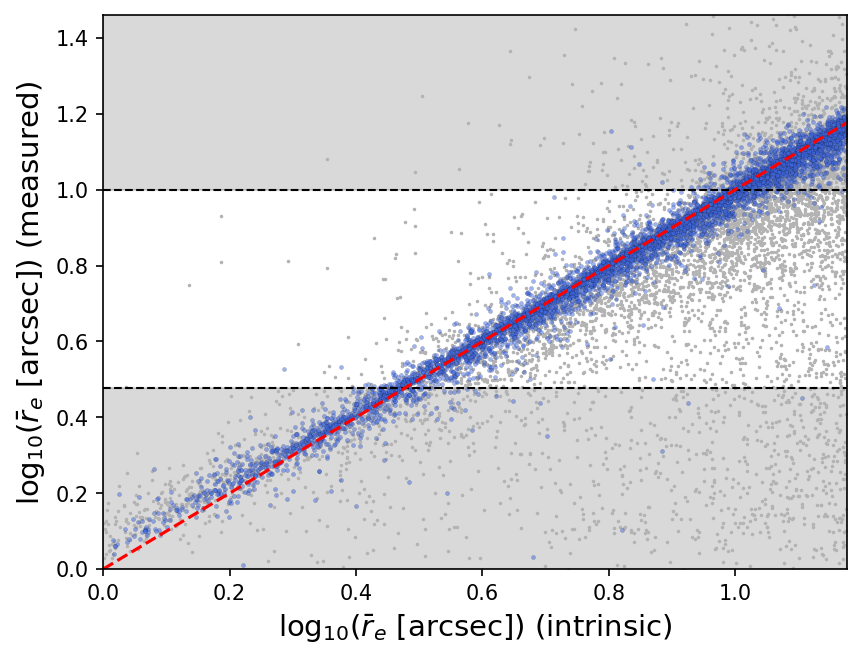}
	\centering
	\caption{The measured circularised effective radii vs. their intrinsic values from our artificial source injections following preselection. The red dashed line is the one-to-one relation. Blue points are those that are predicted by the ANN to have fractional errors less than \MLerrtol. Grey points are those that are rejected based on this criterion. The area that is not shaded corresponds to our selection range in \rec.}
	\label{figure:injectionsSizes}
\end{figure}

\indent While the majority of our recovered injections are measured precisely, there is also a non-negligible population of sources with catastrophic ($>$20\%) error as shown in figure \ref{figure:injectionsSizes}. These measurements have a tendency to be underestimates, which can lead to bias in the analysis. Visual inspection of these artificial sources reveals that many of them have been placed on-top of or in close proximity to real sources in the data which often dominate.

\indent Identifying poor fits automatically is a difficult task but is possible if sufficient information is captured within the source catalogue. In order to combine the high-dimensional data (our catalogue contains \MLnfeatures\ features for each source) to make a prediction of the \rec\ measurement error, we have trained an artificial neural network (ANN) to predict the fractional residual in \rec\ using artificial source injections. We note that a separate, equally-sized set of artificial sources was used to train the network. Further details of the ANN, including its architecture, training and a brief description of the features we used can be found in appendix \ref{appendix:ML}.

\indent The trained ANN allows us to include an additional step in source selection that greatly reduces the number of catastrophically-poor size estimates left in the sample, albeit at the cost of completeness. The aggressiveness of this machine-learned (ML) selection can be tuned to achieve an optimal balance between the number of poor fits and the completeness, modifying \REparam\ accordingly. The effect of this additional selection is displayed in figure \ref{figure:injectionsSizes}. The overall precision in measured sizes is significantly improved within our selection range following the ML selection. We discuss the impact of the ANN selection on our analysis in $\S$\ref{section:annbias}.

\indent A second benefit of training the ANN to recognise extremely bad fits is the removal of misdetections (i.e. non-LSB galaxies) from our sample. While we did not explicitly train the ANN to reject these sources, they are present in the training set because they often coincide with the spatial positions of the artificial sources. Often they are much brighter than the artificial sources and dominate the light. The structural parameter measurements for these misdetections are therefore significantly different from the artificial source, so the ANN learns to reject them. This is discussed further in appendix \ref{appendix:ML}.

\begin{table*}
	\label{table:sourcecat}
	\begin{tabular}{ccccccccccc}
		\hline
		\texttt{ra}       & \texttt{dec}     & \texttt{rec\_arcsec} & \texttt{n\_sersic} & \texttt{mueff\_av} & \texttt{axrat} & \texttt{g\_r} & \texttt{r\_i} & \texttt{weight} & \texttt{is\_red} & \texttt{inside\_gama} \\ \hline
		149.2038 & -1.7671 & 3.3         & 0.97      & 25.3      & 0.78  & 0.57 & 0.38 & 2.27   & True    & False        \\
		148.9471 & -1.8950 & 4.3         & 0.61      & 26.1      & 0.42  & 0.73 & 0.38 & 3.67   & True    & False        \\
		148.4513 & -1.7697 & 3.8         & 1.45      & 24.3      & 0.97  & 0.51 & 0.29 & 1.26   & False   & False        \\
		148.1515 & -1.9437 & 4.0         & 1.06      & 24.5      & 0.57  & 0.35 & 0.15 & 1.26   & False   & False        \\
		150.1394 & -1.6974 & 3.2         & 1.61      & 24.2      & 0.90  & 0.31 & 0.23 & 1.43   & False   & False        \\
		.        & .       & .           & .         & .         & .     & .    & .    & .      & .       & .            \\
		.        & .       & .           & .         & .         & .     & .    & .    & .      & .       & .            \\.        & .       & .           & .         & .         & .     & .    & .    & .      & .       & . \\ \hline
	\end{tabular}
	\caption{Example of the source catalogue provided as an online supplement. The RA, Dec pairs are quoted in degrees. Morphological parameters are given based on the $r$-band S\'ersic fits: \texttt{rec\_arcsec} is the circularised effective radius in arcseconds. \texttt{mueff\_av} is the mean surface brightness within one effective radius, given in magnitudes per square arcsecond. \texttt{weight} corresponds to the inverse of the recovery efficiency. \texttt{is\_red} indicates whether the source belongs to the red LSBG population. \texttt{inside\_gama} indicates whether the source is contained within the overlapping GAMA footprint.}
\end{table*}

\subsubsection{Colour measurements}
\label{section:colours}

\indent In addition to the structural measurements, we also measure the relative brightness of each selected source in all five bands ($grizy$). We measure fluxes in elliptical apertures out to one $r$-band effective radius. We do not correct for PSF effects for these measurements because we expect the corrections to be very small since the PSF is significantly smaller than our lower selection cut in size and the PSF size is similar among the HSC bands. 

\indent Using artificial galaxy injections which account for the PSF variation in different bands, we find that the 1$\sigma$ error in ($g$-$r$) is $<$0.05 mag for the range of colours expected for local galaxies measured by the GAMA survey \citep[\GAMAcolourcat][]{Driver2016}. We do not find evidence for significant systematic bias in our colour measurements between bands.

\begin{figure*}
	\includegraphics[width=\linewidth]{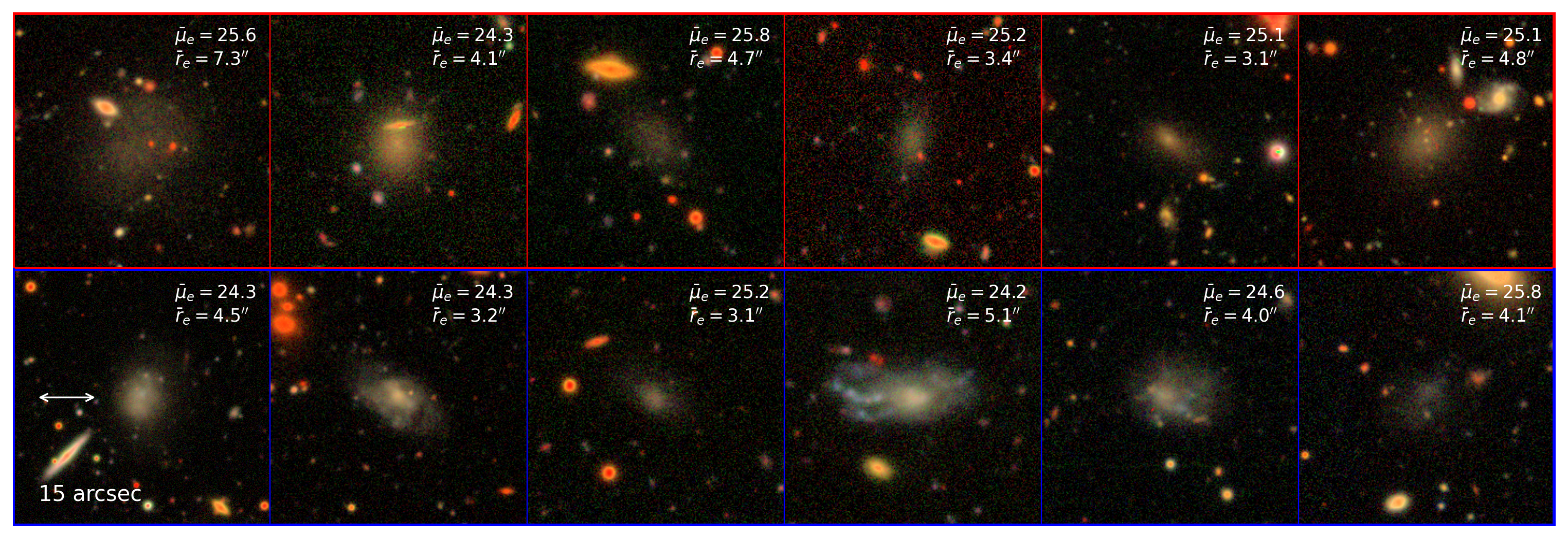}
	\centering
	\caption{Examples of sources in our catalogue following selection and human validation. The upper panels show galaxies in the red population, while the bottom panels show some in the blue population. The cut-outs are 40\arcsec$\times$40\arcsec. The images are created using the HSC $g$, $r$ and $i$ bands and are coloured using the method of \protect\cite{Lupton2004}.}
	\label{figure:sourceplot}
\end{figure*}


\subsubsection{Human validation}

\indent We also performed visual rejection of undesirable sources, defined as those being obviously unassociated with a genuine LSB galaxy, e.g. the outskirts of a bright galaxy's halo or a high redshift galaxy cluster. Three of us (D.P., R.vdB. \& M.H.) used the same software to ``vote'' on which sources to keep; only those unanimously agreed-upon were selected for the final analysis. In no particular order, 80\%, 80\% and 76\% of the \SRCnselnohuman\ selected sources were chosen by the individual voters, with the overall total equalling 73\%, or \SRCnsel\ sources. The consistency between these scores indicates minimal human bias in this selection. This is a marked improvement over the $\sim$50\% of sources rejected visually by \cite{Greco2018}, although the comparison is not completely clean owing to the subjective nature of human validation and intrinsic differences between the two surveys (see also appendix \ref{appendix:greco}). 

\indent While we cannot perform the same human validation for the artificial sources used to calculate the recovery efficiency, we argue that applying this selection to the observed catalogue likely results in only a small change to the overall recovery efficiency. This is because we have tried to only remove sources obviously unassociated with real LSB galaxies, commonly misdetections in the vicinity of brighter galaxies. We are therefore subtracting from the interloping population of misdetections, rather than the actual LSB galaxy population which the recovery efficiency pertains to. Since we find that our main results are not significantly changed by accounting for the recovery efficiency, we disregard this effect as insignificant.


\subsection{The source catalogue}
\label{section:srccat}

Following preselection, we are left with \SRCnpresel\ sources. After applying the full set of selection criteria, \SRCnsel\ remain, \SRCnselgama\ of which are contained within the three overlapping GAMA regions. Note that these numbers include the 27\% of sources rejected by human classification. For reference, the full list of selection criteria are provided below:

\begin{enumerate}
	
	\item Surface brightness:  \SELuaeL$<$\uae$<$\SELuaeU. This targets the LSB regime consistently with other studies \citep{vanderBurg2017, Prole2019b}.
	
	\item Apparent size: \SELrecL$<$\rec$<$\SELrecU. This selects extended sources which are likely to be local galaxies, while removing those which extend over spatial scales similar to that of the background subtraction.
	
	\item S\'ersic index: $n$$<$\SELnU. This is appropriate as LSB galaxies typically have $n$$\sim$1 and helps to remove high redshift early-type interlopers which might have higher indices.
	
	\item Colour: 0$<$($g$-$r$)$<$1. This encapsulates the typical colour range for LSB galaxies and helps to remove high redshift interlopers such as detected clusters of unresolved background galaxies.
	
	\item ML selection: $|\Delta\rec|$$<$\MLerrtol. This statistically selects accurate fits.
	
	\item Axis ratio: We reject sources with \Imfit\ axis ratios exactly equal to 1 because these typically correspond to poor fits.
		
\end{enumerate}

\indent Additionally, we use the improved bright star masks provided as part of HSC-SSP DR2 to mask any sources in the vicinity of stars, as visual inspection of the sources revealed that stellar haloes were a significant source of contamination. We also require individual sources to be separated by \SELuniquedist\ to avoid counting the same source twice in overlapping patches. Several examples of sources in our catalogue are shown in figure \ref{figure:sourceplot}.

\indent We note that following source selection, we remeasured each source using \Imfit\ with the HSC-SSP PSF models. This was done in order to reduce the potential for bias and artificial correlations between S\'ersic parameters that might preferentially affect the smaller of our galaxies. No additional selection was performed after these measurements in order to preserve the recovery efficiency.

\indent An important verification that we are indeed probing a population of local LSB galaxies (rather than being dominated by high-redshift interlopers) can be obtained by comparing the properties of our observed sample with those expected from LSB galaxies. In particular, we find that the range in colour and S\'ersic index of our sources is highly consistent with LSB galaxies from the literature (figure \ref{figure:colourindex}). 

\indent We note that our catalogue is made public as supplementary material, an example is given in table \ref{table:sourcecat}. A comparison between the catalogue produced for this work and the comparable catalogue produced by \cite{Greco2018} is presented in appendix \ref{appendix:greco}.

\subsubsection{A bimodal population}

\begin{figure}
	\includegraphics[width=\linewidth]{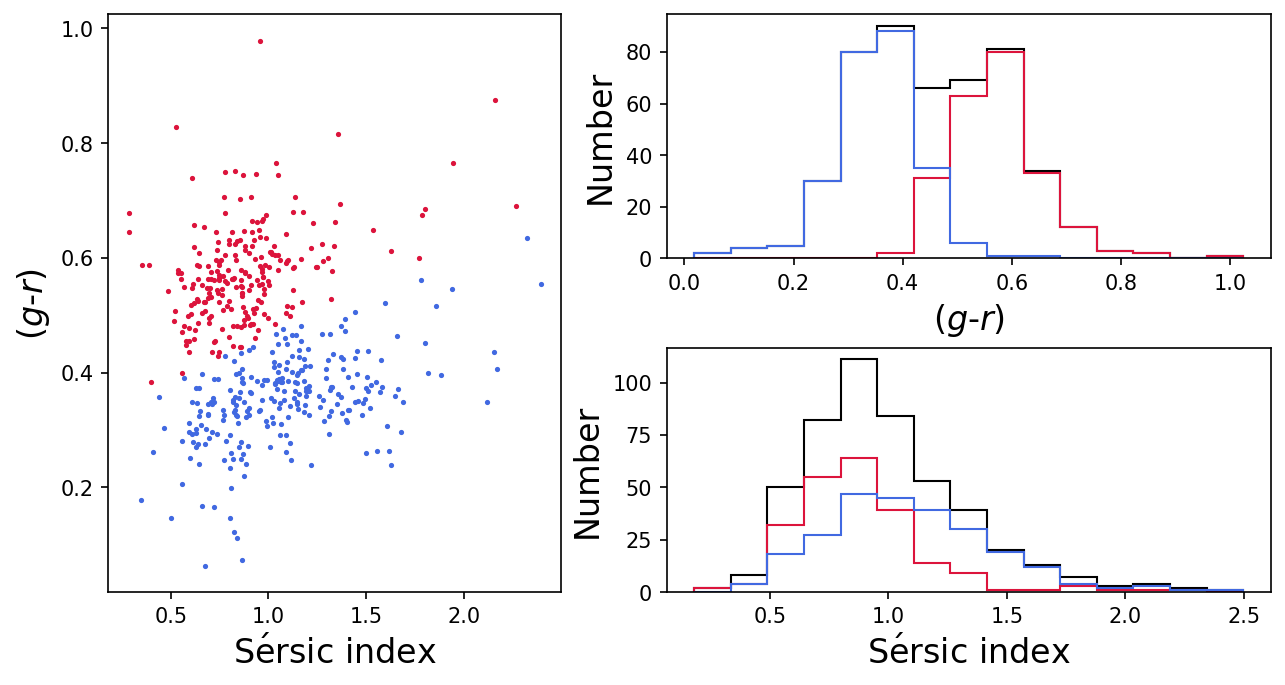}
	\centering
	\caption{Evidence for a bimodal population in colour and S\'ersic index. The colours correspond to a two component $k$-means clustering model. The disparity between the two populations in terms of S\'ersic index only becomes apparent when plotted against colour.}
	\label{figure:colourindex}
\end{figure}

\indent We find evidence for a bimodal population in colour \footnote{although the redder group does not appear red enough to resemble the quiescent UDGs typically found in massive galaxy clusters \citep{Prole2019b}.}, which is more apparent in the colour-S\'ersic index plane; the bluer population tends to have higher indices and therefore more-concentrated profiles. The relative depletion in-between the two populations visible in figure \ref{figure:colourindex} may indicate a fast transition time between the two populations.

\indent The peak value of $n$$\sim$0.7 for the red population agrees well with that found for UDGs inside galaxy clusters by \cite{Roman2017a}, but is slightly smaller than that found by \cite{Koda2015} for those in the Coma cluster (median of $n$=0.91, although their distribution actually peaks at $n$$\sim$0.8). By contrast, the blue population peaks at $n\sim1$, which is the fiducial value for LSB dwarf galaxies in the literature \citep[e.g.][]{Davies1988}.

\begin{figure}
	\includegraphics[width=\linewidth]{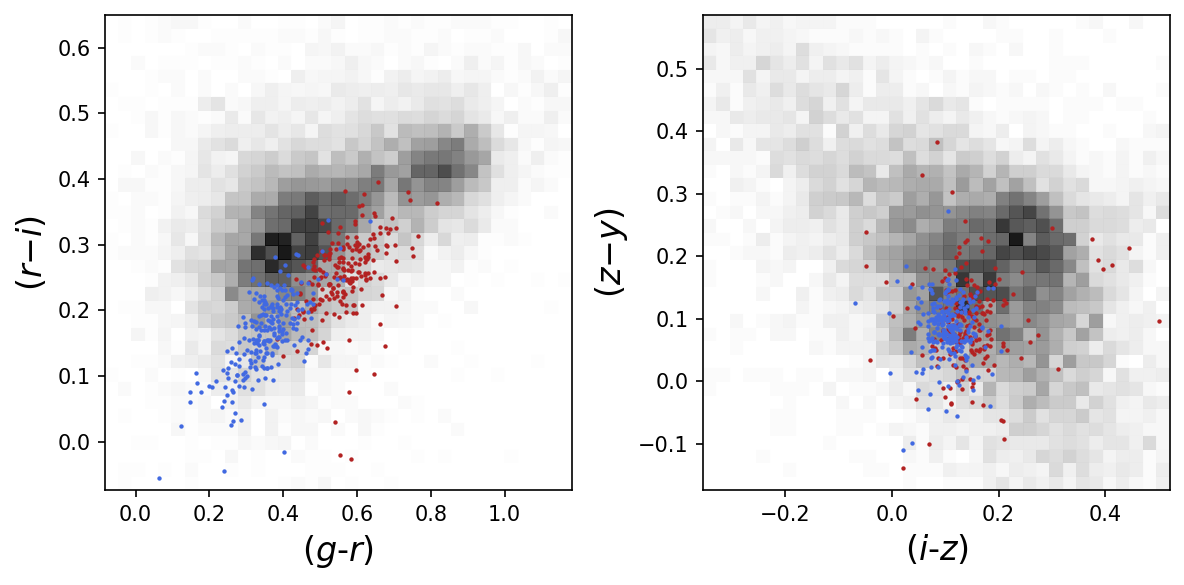}
	\centering
	\caption{Colour-colour diagrams for our LSBGs (coloured red/blue according to their classification) vs. local galaxies measured as part of the GAMA survey (greyscale histogram). Note that we do not account for the small corrections between HSC filters and those used for the GAMA measurements in this comparison.}
	\label{figure:colourcolour}
\end{figure}

\indent Our full sample of \SRCnsel\ selected LSBGs, spread over the \patcharea\ deg$^{2}$ HSC area, is used to define the red / blue populations. \SRCnselb\ of these sources belong to the blue population (\SRCnselbperc\%), whereas \SRCnselr\ (\SRCnselrperc\%) sources belong to the red population. If we only consider the \SRCnselgama\ sources in the overlapping GAMA footprint, there are \SRCnselgamab\ (\SRCnselgamabperc\%) and \SRCnselgamar\ (\SRCnselgamarperc\%) sources in the blue / red populations respectively.

\indent We show the full set of colours for our sources in figure \ref{figure:colourcolour}, compared with local ($z$$<$0.1) galaxies from the GAMA survey (\GAMAcolourcat). The sources are predominantly bluer than their brighter GAMA counterparts, favouring their candidacy for local dwarf galaxies.

\begin{figure}
	\includegraphics[width=\linewidth]{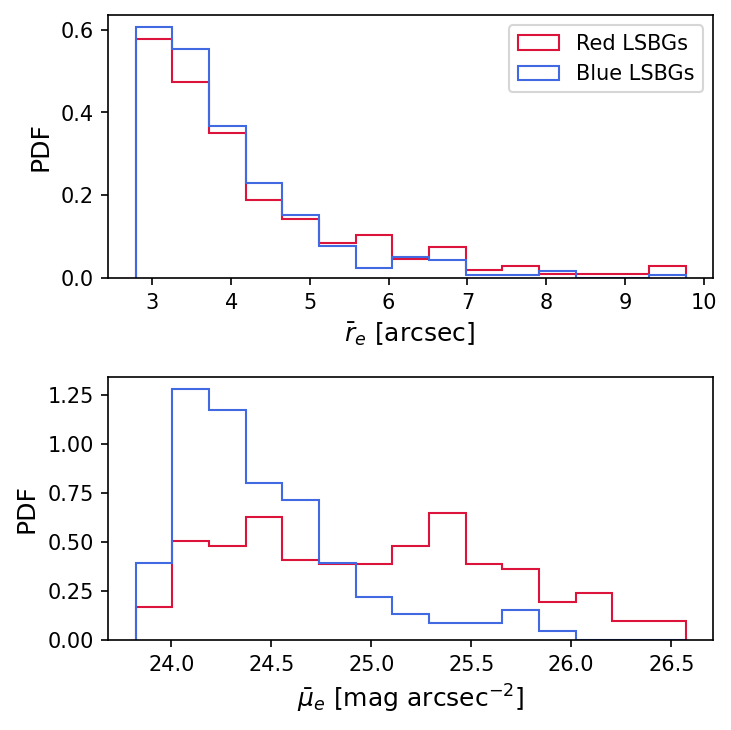}
	\centering
	\caption{Distributions of circularised effective radius (\rec, top) and $r$-band surface brightness averaged within one effective radius (\uae, bottom) for our final sample of candidate LSBGs, plotted according to their population classification. While the distribution of \rec\ is statistically indistinguishable between the two populations, it is clear that blue LSBGs tend to have higher surface brightness.}
	\label{figure:sbsize}
\end{figure}

\indent We find that the distributions of \rec\ (apparent circularised effective radius) and axis ratio are statistically indistinguishable between the two populations (figure \ref{figure:sbsize}). Under the assumption that the physical sizes of LSBGs are distributed similarly, the similarity in apparent sizes between the two populations indicates they are at similar distances.

\indent In figure \ref{figure:sbsize}, we also show that blue LSBGs tend to be brighter than their red counterparts. This is a similar observation to that of \cite{Greco2018}, and is plausibly explained by the presence of young blue stars in blue LSBGs.

\subsection{Crossmatch with GAMA spectroscopic catalogue}

\indent We crossmatched our LSBG catalogue with GAMA spectroscopic sources (\texttt{SpecObjv27}) within 5\arcsec of the central coordinates of the S\'ersic fits and found 24 matches. Since the GAMA spectroscopic survey suffers surface brightness incompleteness in the LSB regime that we probe here, it is no surprise to find that the majority of these matches are at the brighter end of our surface brightness range.

\begin{figure}
	\includegraphics[width=\linewidth]{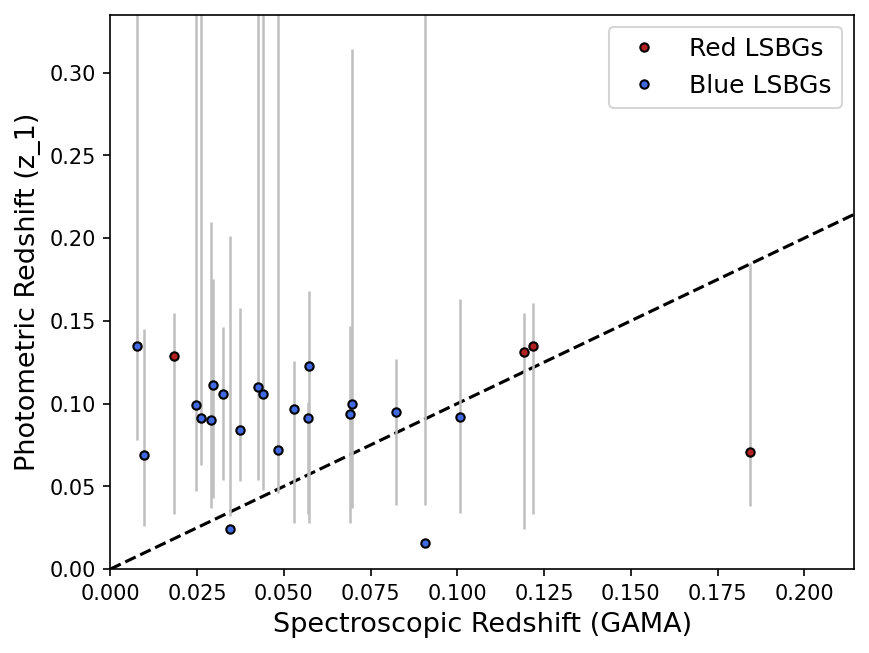}
	\centering
	\caption{Spectroscopic redshifts from the GAMA spectroscopic survey vs. photometric redshifts obtained using \Eazy\ for the crossmatched LSBG sample. The points are colour-coded appropriately for the red and blue LSBG samples. The error-bars on the photometric redshifts enclose the 68\% confidence intervals centred on the median value. In some cases, the maximum-likelihood photometric redshift estimate lies outside of this range (see text for discussion). We note the photometric redshifts are fit between \zphotmin\ and \zphotmax.}
	\label{figure:specphotz}
\end{figure}

\indent We find that essentially all of the blue LSB sources with GAMA matches are below $z$=0.1, whereas the red LSBG matches are all below $z$=0.2. These measurements are displayed in figure \ref{figure:specphotz}. 

\indent We point out that these redshift distributions are not necessarily representative of our full sample owing to the differing completeness functions between our survey and GAMA. As expected, the GAMA cross-matches are systematically brighter than their counterparts without matches. However, we find that the cross-matched sources have a statistically indistinguishable distribution of apparent sizes when compared with the full LSBG sample, suggesting they are at similar distances. 

\subsection{Photometric redshifts}

We use our multi-band HSC $grizy$ measurements (integrated to 1$r_{e}$) to estimate photometric redshifts for our LSBG sample using \Eazy\ \citep{Brammer2008}. We do not expect precise distance estimates from this method; its primary usage is to identify high-redshift interlopers.

\indent We use \Eazy\ in single template combination mode using the \texttt{Pegase13} templates \citep{Grazian2006}. We do not apply a redshift prior and fit redshifts between \zphotmin\ and \zphotmax\ using \texttt{Z\_STEP=0.001} and \texttt{Z\_TYPE=1}. Uncertainties in fluxes are estimated statistically based on our artificial source measurements.

\indent We found that the \texttt{z\_1} redshift estimator (i.e. the redshift of maximal likelihood) in \Eazy\ gave optimal results. This is because other estimators seemed prone to dramatically overestimating the redshifts of the spectroscopic-crossmatched LSB sample, which is particularly undesirable for this analysis. Overall, we find that the photometric redshifts are likely positively biased at low redshift, but are nevertheless accurate to within $\Delta z$$\sim$0.1.

\begin{figure}
	\includegraphics[width=\linewidth]{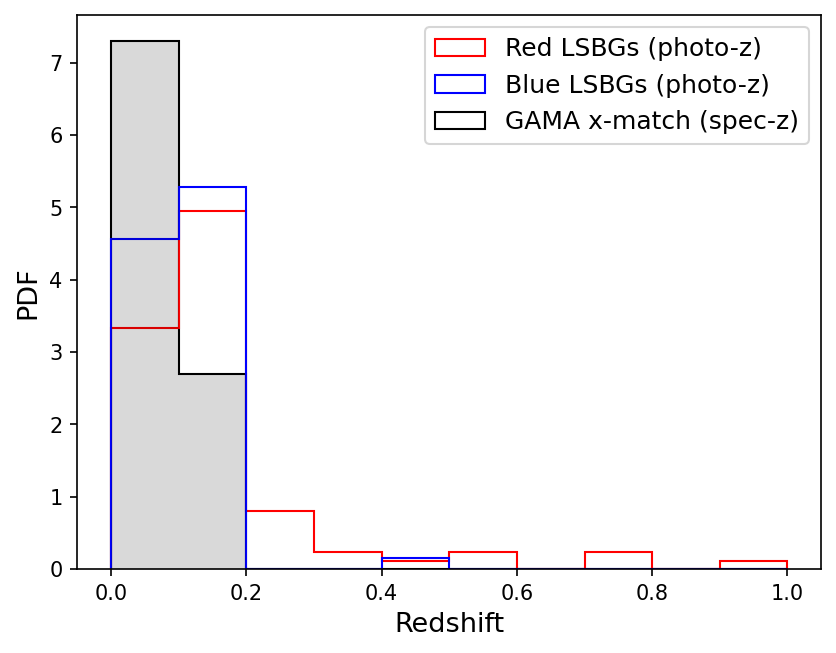}
	\centering
	\caption{Photometric redshifts for the sample of LSBGs contained within the GAMA overlap region, colour-coded appropriately for each of the red and blue LSBG subsamples. The characteristic uncertainty in the photometric redshifts is approximately 0.1. We also show the distribution of spectroscopic redshifts obtained via crossmatching our LSBG catalogue with the GAMA spectroscopic catalogue in black, as described in the text.}
	\label{figure:photzs}
\end{figure}

\indent These photometric redshift estimates are compared with spectroscopic redshifts in figure \ref{figure:specphotz} for the crossmatched sample, whereas the photometric redshift distributions for the full sample are shown in figure \ref{figure:photzs}. We note that this figure only shows the redshift distribution of the LSBGs in the GAMA-overlap region because these are the ones used for the main analysis.

\section{Analysis}
\label{section:analysis}

\indent Our analysis method relies on associating the LSBGs with local galaxies with spectroscopic redshift distance estimates in the GAMA survey spectroscopic catalogue. This approach can be used to statistically estimate the local environment density of the LSBGs and therefore to identify sources that are likely in low-density environments. 

\subsection{Distance arguments}
\label{section:distances}

We present the following arguments that our sample are predominantly at redshifts below 0.2:

\begin{itemize}
	
	\item \textit{Colour}: There are a general lack of galaxies with observer-frame $(g-r)<0.5$ above $z$=0.2 \citep{Driver2016}. Therefore, the blue LSBGs are likely at $z$$<$0.2.
	
	\item \textit{Photometric redshifts}: While our photometric redshift estimates are imprecise, they clearly show that the majority of the LSBGs are at $z$$<$0.2. However, some of the red LSBGs may be at higher redshifts.
	
	\item \textit{Spectroscopic redshifts}: All of the LSBGs with spectroscopic crossmatches from the GAMA survey are at $z$$<$0.2. The rest of the sample have similar apparent sizes, so are likely at similar distances.
	
	\item \textit{Surface brightness and size}: At $z$=0.2, our lower limit cut in effective radius at \SELrecL\ corresponds to a physical size of $\sim$10 kpc. Therefore, if these galaxies are at $z$$>$0.2, they require stellar masses $>$10$^{10}$ M$_{\odot}$ based on empirical galaxy scaling relations \citep{vanderWel2014, Lange2015}. However, at $z$=0.2, one expects $<$1 mag of surface brightness dimming, assuming the (1+$z$)$^{4}$ relation. Such massive galaxies are typically much too bright \citep{Wright2017} to then satisfy our LSB criterion of \uae$\geq$\SELuaeL\ mag arcsec$^{-2}$. Therefore, the LSBGs are likely at $z$$<$0.2 because they are too LSB for their apparent sizes.
	
		
\end{itemize}

Further, we present the following arguments that most of our sample (particularly blue galaxies) are likely at redshifts below 0.1:

\begin{itemize}
	
	\item \textit{Spectroscopic redshifts}: The spectroscopic redshifts from the GAMA cross match clearly show that blue LSBGs are at $z$$<$0.1.
	
	\item \textit{Empirical modelling}: \cite{Prole2019b} modelled a population of UDGs using empirical scaling relations. Accounting for their recovery efficiency, which is comparable to that in this study, they found most detectable UDGs are at $z$$<$0.1.
	
	\item \textit{Photometric redshift precision}: As discussed, our photometric redshift estimates are accurate to within $\Delta z$=0.1. If there were significant numbers of sources actually at $z$$>$0.1, we would expect to measure more photometric redshifts beyond $z$=0.2 simply due to measurement error. Again, this is a stronger argument for the blue LSBG sample based on figure \ref{figure:photzs}.
	
\end{itemize}

To summarise the above, we strongly argue that the majority of our LSBGs are at $z$$<$0.2, but the photometric redshifts suggest some of the red LSBGs might be at higher redshifts. We also argue that the population of blue LSBGs are predominantly within $z$$<$0.1.

\subsection{Spectroscopic galaxy association}
\label{section:knn}

\indent We use the mean distance to the nearest $k$ GAMA sources, \dk, to estimate environmental density for each LSBG. Because this method relies on GAMA spectroscopic sources, we are limited to using the subsample of \SRCnselgama\ LSBGs within the overlapping GAMA footprint. This is similar to the analysis undertaken by \cite{Roman2017a}, but for a larger sample and over a much wider area. 

\indent We use the GAMA \GAMAstellarmasses\ catalogue \citep{Taylor2011} to select spectroscopic sources because it allows us to impose cuts in stellar mass. We use two sets of redshift / stellar mass cuts: One with $z$$<$0.1, $M_{*}$$>$10$^{9}$M$_{\odot}$, and the other with $z$$<$0.2, $M_{*}$$>$10$^{10}$M$_{\odot}$. The lower-limit stellar masses are imposed to avoid redshift-incompleteness effects \citep[e.g.][]{Wright2017}, whereas the redshift cuts are justified by the distance arguments presented in \ref{section:distances}. 

\indent There are many ways to define the nearest neighbours, but here we focus on two (and note here they both give consistent results):

\begin{enumerate}
	
	\item The simplest approach is to find the nearest neighbours and quote transverse distances in angular units. The obvious drawback is that since we are probing a conic volume element, there are more spectroscopic sources towards higher redshifts. In order to partially mitigate this effect, we weight the average distances using the inverse distance cubed. 
	
	\item For the second method, we calculate the co-moving transverse distances (in physical units) between each LSBG and all spectroscopic sources, assuming they are at the same distances. These traverse distances are then used to define the nearest neighbours and we quote transverse distances in physical units.
	
\end{enumerate}

\indent Clearly, both methods of estimating density are error-prone because of projection effects within the volume element. We therefore anticipate that many of the \dk\ measurements will not be physically meaningful, but will be instead drawn from a spatially randomised component that does not spatially correlate with the spectroscopic galaxies. The complement of this spatially randomised component are the LSBGs that are actually associated with local spectroscopic galaxies. We term these as the randomised and associated components respectively. The mixing fraction between the randomised and associated components may be different for the red and blue LSBG populations.

\indent We estimate the \dk\ distribution corresponding to spatially randomised sources by randomly sampling several thousand RA/Dec pairs from our survey footprint and measuring \dk\ at the randomised locations. This allows us to quantify the probability density function (PDF) of \dk\ for the randomised component, which can be compared with the actual measured \dk\ distribution.

\subsection{Density measurements}
\label{section:density}

\begin{figure*}
	\includegraphics[width=\linewidth]{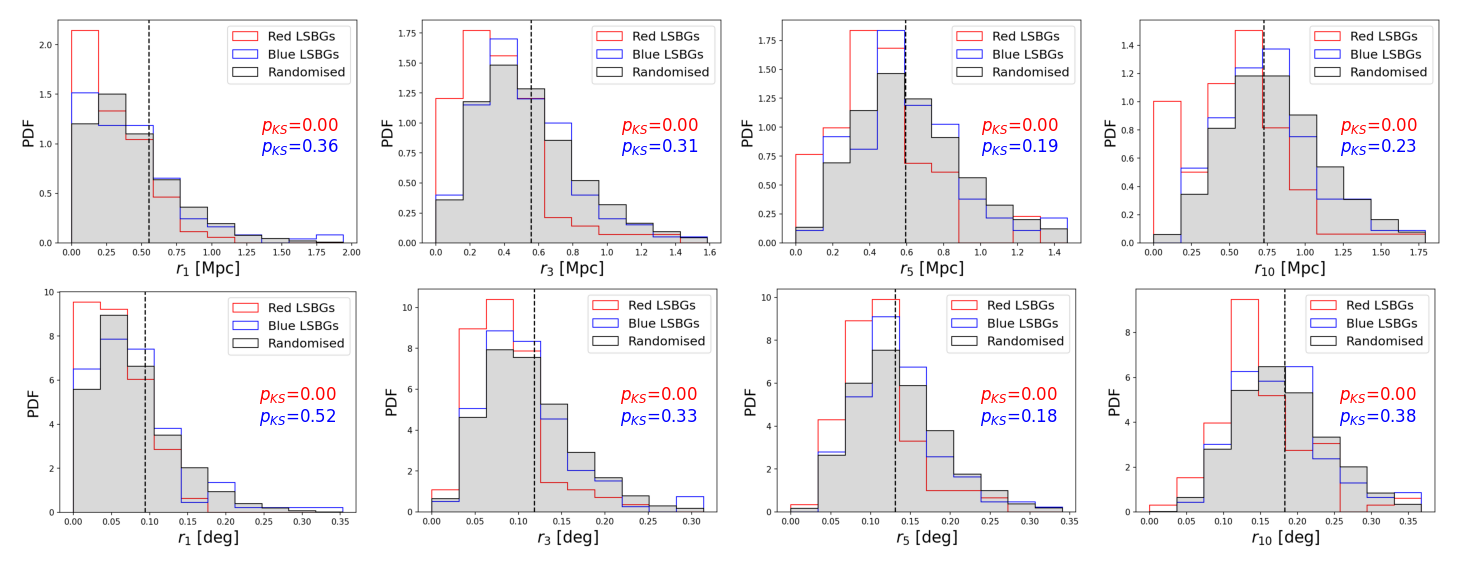}
	\centering
	\caption{Distributions of the environmental density estimator \dk. The top panels show the results using the physical nearest neighbours approach, whereas the bottom show the results from using angular coordinates to define nearest neighbours. The grey shaded histogram corresponds to the \dk\ distribution of spatially randomised sources. The blue and red histograms correspond to the blue and red LSBG samples respectively. The vertical dashed line indicates the outside groups condition. In all cases, the KS statistic $p$-values indicate the blue LSBGs are statistically consistent with being drawn from the same parent distribution as the spatially randomised population, whereas the red LSBGs are not.}
	\label{figure:dk}
\end{figure*}

We test five different values of $k$: \{1, 3, 5, 7, 10\}. When $k$=1, the density is defined as the distance to the nearest neighbour. Higher values of $k$ may give a more robust density estimate because distances are averaged over more galaxies, but are not as sensitive to small over-densities. We show in $\S$\ref{section:qf} that our main results are not sensitive to the choice of $k$. We show the \dk\ distributions for the red LSBG, blue LSBG and spatially randomised populations in figure \ref{figure:dk}.

\indent Independently from the choice of $k$ or upper-limit cut in spectroscopic redshift, we find that the \dk\ distribution measured for blue LSBGs is statistically consistent with being completely spatially randomised. This is verified using two-sample Kolmogorov–Smirnov (KS) tests, in which we are unable to reject the null-hypothesis that the blue LSBGs and spatially randomised \dk\ distributions are drawn from the same parent distribution (typical $p$-values are $>$0.2). The fact that there is no dependence on this observation with $k$ leads us to believe that the blue LSBGs are indeed predominantly unassociated with local structure over-densities.

\indent However, we find that the red LSBG sample is not consistent with being completely spatially randomised, with KS $p$-values $<0.01$ when considering spectroscopic galaxies at $z$$<$0.1. We note that the significance of this detection of associated structure is not as high when considering spectroscopic galaxies with $z<0.2$, where we are unable to show the distribution is not random at similar significance (KS $p$-values $<$0.1). We therefore proceed with the cut at $z=0.1$ for our main result. This is discussed further in \ref{section:zdep}. We note that the KS tests give the same results when sources with photometric redshifts significantly (1$\sigma$) outside of the selection range are used for both red and blue LSBGs.

 
\subsection{Density measurements of GAMA groups}
\label{section:groups}

It is important to relate the \dk\ statistic to something more physically meaningful in order to interpret our results. To do this, we make use of the publicly available GAMA group catalogue \citep[\texttt{GroupFinding v10}, ][]{Robotham2011}, which covers the G15 field, approximately one-third of the footprint considered here.

\indent Groups are selected according to the redshift cut we use for the spectroscopic galaxy associations. In order to select robust groups, we require more than 5 friends-of-friends members for each group. We estimate total mass ($M_{200}$) from $r$-band luminosities using the empirical scaling relations from \cite{Viola2015}, from which we also derive $r_{200}$. We select groups with $M_{200}$$>$$10^{12}$M$_{\odot}$. Overall, we select 105 groups at $z$$<$0.1 and 502 groups at $z$$<$0.2.

\indent In order to find the range of \dk\ associated with these groups, we randomly sample RA/Dec pairs within $r_{200}$ for each group, measuring corresponding \dk\ values.  We define LSBG galaxies as ``outside groups'' if they have an \dk\ measurement greater than the 95th percentile of the \dk\ distribution associated with all slected groups, quantified using the above method. This criterion is indicated in figure \ref{figure:dk}, which corresponds to the dividing line between galaxies inside and outside of groups.

\subsection{Quiescent fraction of isolated LSBGs}
\label{section:qf}

As discussed previously, the blue LSBG population is consistent with being completely dissociated with local structure over-densities, yet all are likely to be within the spectroscopic galaxy redshift selection limits. While we have shown that many of the red LSBGs are associated with high-density environments, there is a non-negligible fraction of these galaxies that are plausibly unassociated with local structure over-densities. There are two explanations for this: either they are part of a genuine population of spatially randomised local red LSBGs, or they are outside the redshift range used to select spectroscopic sources used for density estimation.



\begin{figure}
	\includegraphics[width=\linewidth]{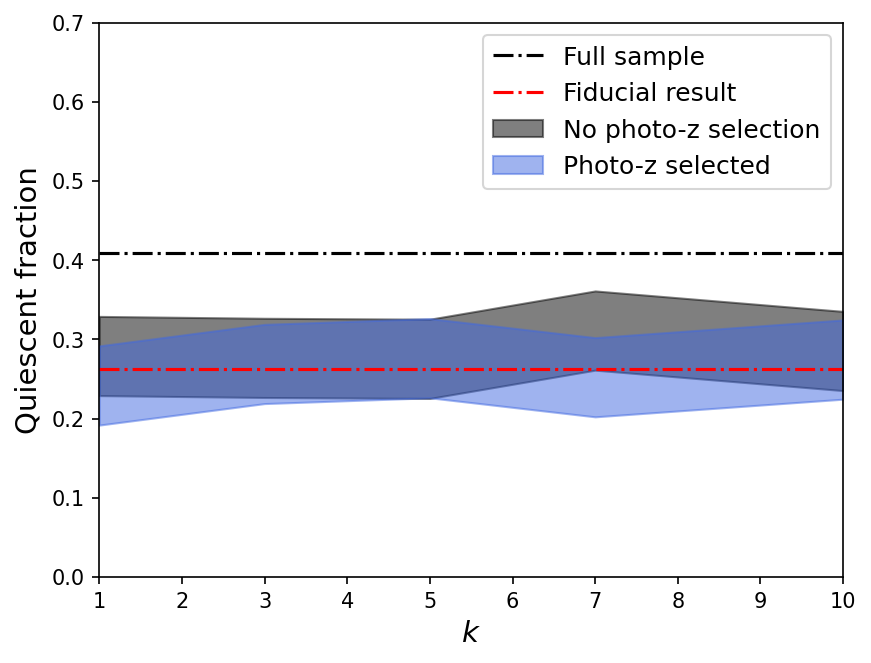}
	\centering
	\caption{Quiescent fraction of isolated galaxies as a function of $k$. The black dashed line corresponds to the full sample of \SRCnselgama\ irrespective of their environment density estimates. The grey region corresponds to the upper-limit estimate from LSBGs outside of the galaxy group \dk\ threshold without applying the photometric redshift selection. The blue region corresponds to our best estimate of the quiescent fraction, combining the group threshold with photometric redshift selection. The dashed red line is our fiducial result. See text.}
	\label{figure:qf}
\end{figure}

We obtain and estimate of the quiescent fraction of ``isolated'' LSBGs by considering only those galaxies that are likely to be outside of groups, as discussed in $\S$\ref{section:groups}. These results are shown in figure \ref{figure:qf}. However, this estimate is an upper-limit because some of the red LSBG population are feasibly beyond the spectroscopic redshift cut and should not be included in the calculation. This upper-limit estimate is approximately \RESqfupper. 

\indent It is possible to obtain an improved upper-limit estimate of the quiescent fraction of isolated LSBGs by removing sources with photometric redshifts significantly (1$\sigma$) outside of the spectroscopic redshift range. However, the photometric redshifts are highly uncertain, meaning that we can only reject a small fraction (10\% for $z$$<$0.1 and 1\% for $z$$<$0.2) of sources using this method. Nevertheless, we find that the quiescent fraction slightly drops as a result of using them, as expected. This is also shown in figure \ref{figure:qf}.

\indent We additionally note that all of our results are corrected for incompleteness by weighting individual observations with the inverse of their recovery efficiencies. Given the lack of dependence on our result with $k$, we take the mean quiescent fraction over $k$ values, obtaining \RESqf. The uncertainty in this value is driven by sampling uncertainty in the analysis as well the scatter with $k$. This value is derived using the density estimates in physical units, which we found to place tighter constraints on the quiescent fraction. 

\subsection{Dependence on spectroscopic redshift cut}
\label{section:zdep}

We considered two cuts in spectroscopic redshift for GAMA sources which were used to estimate densities: One at $z$=0.1 and the other at $z$=0.2. These are motivated by the distance arguments presented in $\S$\ref{section:distances}. At $z$=\{0.1, 0.2\}, we selected GAMA sources with $M_{*}$$>$$10^{\{9,\ 10\}}$M$_{\odot}$ to avoid incompleteness of spectroscopic sources as a function of redshift.

\indent When using $z$=0.2, we obtain a weaker constraint on the upper-limit quiescent fraction compared to the $z$=0.1 case. This is because there are a lot more spectroscopic sources in the $z$=0.2 case (38787 vs 12791) which increase the amount of background contamination in the density estimates. The net effect of this is to increase the maximum plausible amount of spatially randomised red LSBGs (the blue ones are always consistent with randomisation) and therefore to artificially increase the upper-limit quiescent fraction measurement by around 10\%.

\indent Even though we expect a certain amount of our LSBGs to be at redshifts higher than 0.1, because it gives a lower value for the upper-limit of the quiescent fraction of isolated LSBGs, we take that as our strongest constraint. We account for this effect by using our photometric redshift estimates, which we show work to lower the quiescent fraction estimate.

\subsection{Does the ANN-based selection unfairly target cluster/group members?}
\label{section:annbias}

It is important to address whether the ANN selection may be affecting our result by unfairly removing large LSB galaxies in dense environments. We demonstrate this by showing the rejected sources are not particularly associated with galaxy groups and clusters by exploiting the publicly available GAMA group catalogue \texttt{GroupFindingv10} \citep{Robotham2011}. We note that this test is restricted to the G15 GAMA region covered by the public group catalogue. We select groups with $z$$<$0.1 and $>$5 friends-of-friends members. 

\indent We measure the distances to the nearest GAMA group for each galaxy in our G15 subsample (approximately a third of our sources). We compare the distributions of these distances for galaxies that were rejected by the ANN with those that were kept using a two-sample KS test. We find that the two samples are statistically consistent with being drawn from the same population and that there is therefore no significant tendency for group/cluster members to be preferentially removed from the analysis.

\indent Additional evidence that this is not seriously impacting our result comes from visual inspection of the rejected sources (see figure \ref{figure:sourceplotrej} for some examples), which indicate that the rejected sources are typically associated with the LSB outskirts of brighter objects, which based on their red colours may be at relatively high redshifts.

\section{Discussion}
\label{section:discussion}

\indent There is not much discussion of quiescent LSBGs in low-density environments amongst the recent literature, despite huge advances in observational techniques, particularly in wide-field optical imaging surveys. This is because they are reasonably difficult to identify; their red colours mean they cannot easily be distinguished from background objects and no spectroscopic survey is yet deep enough to pick them out spectroscopically. 

\indent A handful of genuine isolated and quiescent LSBGs have been recently identified \citep{Delgado2016, Roman2019}. The latter of these studies relied upon a distance estimate to the galaxy based on its globular cluster (GC) luminosity function. Such a technique, while useful for LSBGs that exhibit clear GC populations, cannot be expected to provide a census of quiescent field LSBGs which may not harbour such GC systems. This technique is also limited by the detectability of GCs, which is not currently feasible for distances far outside the local volume.

\indent Surface brightness fluctuation distances \citep[e.g.][]{Carlsten2019, Greco2020} may provide precise estimates for these sources out to $\sim$100 Mpc ($z$=0.02), but it is possible their relative rarity will present challenges. Alternatively, observations of supernovae associated with LSBGs may provide reliable distance estimates in future studies \citep{Sedgwick2019}, but such surveys will naturally favour star-forming systems where the supernovae rate is higher. It seems likely that until a spectroscopic survey specifically designed to target extended LSB objects materialises, the community will be forced to contend with photometric redshifts which often lack precision, particularly in the local Universe where LSBGs are more detectable.

\indent In this work, we have sidestepped the difficulties of obtaining individual distance measurements by taking a statistical approach based on apparent proximity to spectroscopic sources. Such methodology may be extremely valuable given upcoming wide spectroscopic surveys like the Taipan galaxy survey \citep{daCunha2017}, which will provide spectroscopic redshifts for over a million bright sources over the Southern hemisphere, and 4MOST \citep{deJong2019}. The kind of analysis we have performed here can be easily scaled to this area and provides yet another synergy between deep sky imaging surveys and comprehensive spectroscopic surveys. 

\indent The main result of this study has been the measurement of the quiescent fraction of LSBGs in the field, which we find to be approximately \RESqf, with an upper limit of approximately \RESqfupper. This result implies that secular evolution is capable of producing a reasonable fraction of quiescent LSBGs, potentially from supernovae feedback, which is known to both produce diffuse LSB  systems \citep{DiCintio2017} and quench star formation \citep{Hayward2017}. This result appears to contradict the previous result of \cite{Prole2019b}, who found no evidence for a significant population of red UDGs in low-density environments, but we point out that the present survey has probed lower surface brightness levels than in that study; since red LSBGs are fainter than their blue counterparts \citep[see also][]{Greco2018, Tanoglidis2020}, the fraction of red LSBs is larger here. Indeed, it is possible that as surveys become deeper, greater numbers of quiescent field LSBGs will be unearthed.


\subsection{Structural parameters}

\indent We do not find evidence for an environmentally-dependent axis ratio distribution. However, this does not mean that we can exclude a trend. Indeed, \cite{Cardona-Barrero2020}, find that approximately half of field UDGs are expected to have rotationally supported stellar disks. We do note that our red LSB  galaxy sample typically have lower S\'ersic indices (and are therefore less concentrated), implying there may be some environmental dependency that we have not detected with significance in our main analysis. This is consistent with predictions of tidal heating/stirring, which can produce changes in morphology and turn disky dwarf galaxies into spheroidals \citep[e.g.][]{Mayer2010}, which are also known to be gas-poor \citep{Gallagher1994} and therefore quiescent, also consistent with the reddening as a function of increasing environmental density that we observe here. 

\indent We additionally find similar distributions in apparent size among the red and blue populations, which indicates similar (to first-order) physical size distributions between the two populations assuming they have a similar redshift distribution, as indicated by the photometric redshift estimates. Similar size distributions are expected theoretically \citep{Amorisco2016}.

\subsection{Comparison with LSB galaxies in DES}

\indent In this section, we compare this study with the recent study of LSB galaxies from \cite{Tanoglidis2020}, based on data from the Dark Energy Survey \citep[DES,][]{Abbott2018}. Their catalogue contains 20,977 sources, making it much larger than our sample.

\indent Both studies report a significant bimodal population in terms of colour. However, \cite{Tanoglidis2020} did not find evidence for a bimodal separation in S\'ersic index. We suggest this is because they define their red and blue populations differently to this study. Whereas they have modelled their red and blue populations with a two-component Gaussian mixture in one-dimensional colour space, we perform a two-dimensional clustering analysis in combined S\'ersic index-colour space, which clearly shows a bimodal population (figure \ref{figure:colourindex}).

\indent A second possibility to explain this difference may be that the catalogue of LSB galaxies from \cite{Tanoglidis2020} is compromised by high-redshift interlopers. These are expected to mainly contaminate the red population and be composed of redshift-dimmed late-type galaxies at intermediate redshifts \citep[][]{Prole2019b}. Given their lower cut in effective radius (2.5$\arcsec$) and lack of an upper-limit cut in S\'ersic index, it is possible that the level of contamination from these objects is somewhat higher than in our catalogue. Since these interlopers are typically more massive than the genuine LSB galaxies, they might be expected to have systematically larger S\'ersic indices \citep[e.g.][]{Danieli2019}.

\indent Both studies agree that redder galaxies are associated with denser environments, confirming results from previous studies \citep[e.g.][]{Roman2017, Prole2019b}. Both studies also find that the distribution of angular sizes for red and blue populations are similar, which implies similar physical size distributions (assuming identical redshift distributions between the two populations), although it is not clear to what extent this is a projection effect.

\subsection{Comparison with cosmological simulations}

\indent Recent work using the New Horizon cosmological simulation has provided further theoretical credence to the existence of large low surface brightness galaxies in low density environments \citep{Jackson2020}. One prediction from this study is that LSB dwarf galaxies born in high density environments become more diffuse than their counterparts in lower-density environments because of significant tidal perturbation. This is supported by our observations that red LSBGs are associated with group/cluster environments $\textit{and}$ tend to be fainter than bluer LSBGs, which are disassociated from large scale structure.

\indent The recent study of \cite{Dickey2020} used three hydrodynamical cosmological simulations (EAGLE, Illustris-TNG, and SIMBA) to constrain the quiescent fraction of isolated LSBGs expected from theory. Their results indicate quiescent fractions approximately between 0.05 and 0.35 (depending on stellar mass and which simulation was used), consistent with observational results from this study. However, their own observational comparison concluded that no quiescent LSBGs exist in isolation. While the results from our study are consistent with this idea, several key examples \citep{Delgado2016, Roman2019} have already proven the existence of isolated quiescent galaxies and so the true quiescent fraction is unlikely to be zero in the field. We suggest that the observational discrepancy can be explained by their observational dataset being shallower than ours, compounded by the fact that quiescent LSBGs are systematically fainter in surface brightness than blue ones.


\section{Conclusions}
\label{section:conclusions}

In this work, we have used \patcharea\ square degrees of deep optical HSC-SSP data to create a new catalogue of \SRCnsel\ LSB galaxy candidates that are likely genuine local ($z$$<$0.2) LSB dwarf galaxies based on their apparent sizes, colours and morphologies and a variety of other distance arguments. We found evidence for a bimodal population on the colour-S\'ersic index plane, with bluer galaxies tending to have higher S\'ersic indices.

\indent We estimated environmental density for a subsample of \SRCnselgama\ LSBGs using a nearest spectroscopic neighbours approach using spectroscopic sources from the GAMA survey. We found that the population of blue LSBGs were consistent with spatial randomisation independently from the hyper-parameters of the analysis, implying they predominantly exist in isolation. The density measurements of the red LSBG population were found to be statistically distinct from a spatially randomised population. 

\indent We calibrated the density estimates on GAMA galaxy groups in order to identify a criterion to identify LSBGs likely to be outside groups. We used photometric redshifts to remove high redshift outliers, which particularly contaminate the red LSBG population. Combining this information, we estimate \RESqf\ of isolated LSBGs are quiescent, with an upper-limit of approximately \RESqfupper\ obtained when no photometric redshift selection was applied. It is possible the quiescent fraction of LSBGs will increase as surveys become deeper, owing to the fact that quiescent LSBGs are fainter in surface brightness. Even so, our results are quantitatively consistent with modern cosmological simulations \citep{Dickey2020}.

\indent Over the next decade, the kind of analysis we have performed here can be significantly improved and expanded upon by combining the wealth of LSB sources anticipated to be found in upcoming deep sky surveys with new wide-field spectroscopic galaxy surveys like Taipan, and has great potential to shed new light on the formation and evolution of LSB galaxies like never before.\newline

\textit{The data underlying this article are available in its online supplementary material. Additional data will be shared on reasonable request to the corresponding author.}


\section{Acknowledgements}

\noindent We are extremely grateful to Yoshihiko Yamada and Masayuki Tanaka for providing technical information regarding the HSC-SSP data. \newline

\noindent D.P. and L.S. acknowledge funding from an Australian Research Council Discovery Program grant DP190102448. \newline

\noindent This research was undertaken with the assistance of resources from the National Computational Infrastructure (NCI Australia), an NCRIS enabled capability supported by the Australian Government. \newline

\noindent The Hyper Suprime-Cam (HSC) collaboration includes the astronomical communities of Japan and Taiwan, and Princeton University. The HSC instrumentation and software were developed by the National Astronomical Observatory of Japan (NAOJ), the Kavli Institute for the Physics and Mathematics of the Universe (Kavli IPMU), the University of Tokyo, the High Energy Accelerator Research Organization (KEK), the Academia Sinica Institute for Astronomy and Astrophysics in Taiwan (ASIAA), and Princeton University. Funding was contributed by the FIRST program from the Japanese Cabinet Office, the Ministry of Education, Culture, Sports, Science and Technology (MEXT), the Japan Society for the Promotion of Science (JSPS), Japan Science and Technology Agency (JST), the Toray Science Foundation, NAOJ, Kavli IPMU, KEK, ASIAA, and Princeton University. \newline

\noindent This paper makes use of software developed for the Large Synoptic Survey Telescope. We thank the LSST Project for making their code available as free software at  http://dm.lsst.org. \newline

\noindent This paper is based [in part] on data collected at the Subaru Telescope and retrieved from the HSC data archive system, which is operated by Subaru Telescope and Astronomy Data Center (ADC) at National Astronomical Observatory of Japan. Data analysis was in part carried out with the cooperation of Center for Computational Astrophysics (CfCA), National Astronomical Observatory of Japan. \newline

\noindent GAMA is a joint European-Australasian project based around a spectroscopic campaign using the Anglo-Australian Telescope. The GAMA input catalogue is based on data taken from the Sloan Digital Sky Survey and the UKIRT Infrared Deep Sky Survey. Complementary imaging of the GAMA regions is being obtained by a number of independent survey programmes including GALEX MIS, VST KiDS, VISTA VIKING, WISE, Herschel-ATLAS, GMRT and ASKAP providing UV to radio coverage. GAMA is funded by the STFC (UK), the ARC (Australia), the AAO, and the participating institutions. The GAMA website is http://www.gama-survey.org/.\newline

\noindent This research made use of Astropy,\footnote{http://www.astropy.org} a community-developed core Python package for Astronomy \citep{astropy2013, astropy2018}.\newline

\noindent This research has made use of the VizieR catalogue access tool, CDS, Strasbourg, France (DOI : 10.26093/cds/vizier). The original description of the VizieR service was published in A\&AS 143, 23.\newline

\noindent This research has made use of NASA's Astrophysics Data System.



\bibliographystyle{mnras}
\bibliography{library.bib}

\appendix

\section{Artificial Neural Network}
\label{appendix:ML}

We trained an ANN to identify anomalously poor \Imfit\ results by making it predict the absolute log-error in the recovered values of \rec\ using a catalogue of \SynthNumber\ artificial galaxy injections. The network was implemented using the Python module \texttt{keras}. The inputs to the network were \MLnfeatures\ features extracted from the statistics of pixels in the vicinity of the detections, specifically:

\begin{enumerate}
	\item Segment statistics: \texttt{R50}, \texttt{mag}, \texttt{q}, \texttt{I50}, \texttt{I50av}, \texttt{SB\_N50} (\DeepScan) and \texttt{R\_e}, \texttt{mu\_max}, \texttt{mu\_median}, \texttt{R\_fwhm} (\MTObjects).
		
	\item \Imfit\ results: \texttt{n\_fit}, \texttt{mag\_fit}, \texttt{q\_fit}, \texttt{rec\_fit}, \texttt{uae\_fit}, \texttt{sky\_I0\_fit}, \texttt{sky\_dIdx\_fit}, \texttt{sky\_dIdy\_fit}.

	\item Additional fit statistics: \texttt{sx\_fit}, \texttt{sy\_fit} (the x and y sizes of the images used for the fitting, in pixels), \texttt{resstd\_fit} (the standard deviation of the unmasked residuals)
	\texttt{modelfluxfrac} (the total flux of the model divided by that of the data), \texttt{offset\_fit} (the positional offset between the flux-weighted segment centroid and the \Imfit\ model centroid), \texttt{maskedfluxfrac} (the fraction of the cutout flux that was masked), \texttt{maskedareafrac} (the fractional cutout area that was masked), \texttt{maskedfluxfrac\_model} (the fraction of the model flux that was masked).
\end{enumerate}

There is clearly some redundancy in these features but our strategy was to incorporate as much information as possible into the training. Each feature was normalised by subtracting its mean and dividing by the standard deviation in the training data. The same normalisation technique was applied to the training labels (logarithmic absolute errors in \rec).

\indent The ANN architecture consisted of five fully-connected layers with 128 (tanh activation), 64 (rectilinear unit activation), 32 (rectilinear unit activation), 16 (tanh activation) \& 1 (linear activation) neurons consecutively. A 1\% dropout layer was used after each layer except the final two in order to reduce over-fitting. The model was trained to minimise the mean squared error using the \texttt{Adam} \citep{Kingma2014} optimiser with a batch size of 256 over 100 epochs. Validation was done using an entirely different set of artificial sources (e.g. blue points in figure \ref{figure:injectionsSizes}), which were also used to estimate the recovery efficiency (figure \ref{figure:RE}).

\begin{figure*}
	\includegraphics[width=\linewidth]{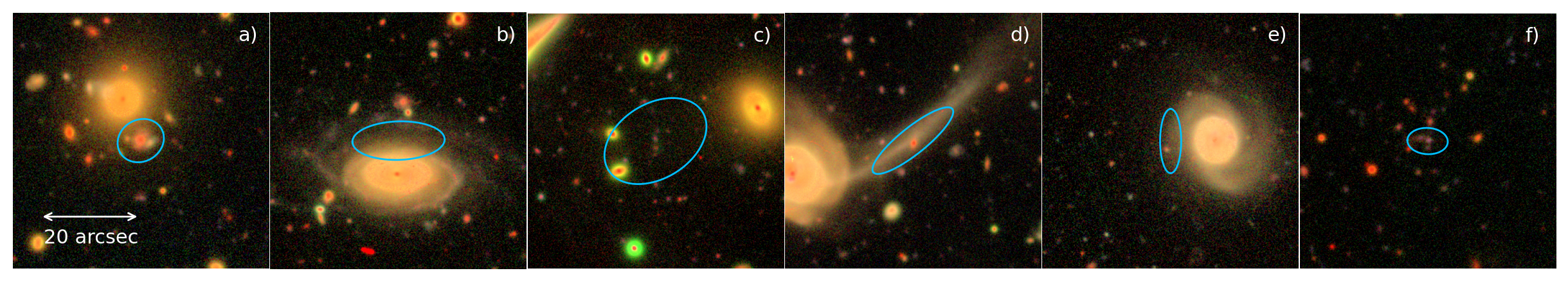}
	\label{figure:sourceplotrej}
	\centering
	\caption{Examples of sources rejected by the ANN for being likely poor fits, that would have otherwise satisfied the selection criteria. The 1$r_{e}$ ellipses are shown in blue. Many of these rejected sources correspond to improperly de-blended haloes of bright sources, tidal features and background galaxy clusters. While the ANN was not trained to specifically identify such cases, the training data encapsulated enough information to identify several of these cases.}
\end{figure*}

\indent We chose to reject sources with an absolute error greater than \MLerrtol. Some examples of rejected sources are shown in figure \ref{figure:sourceplotrej}. Even though the ANN was trained on artificial galaxy injections, it learned to identify misdetections such as outskirts of bright galaxies. This is likely because the injections were placed at randomised positions in the data, so many of them were placed on top of much brighter sources which subsequently skewed the \Imfit\ results.

\section{Comparison with Greco+18 catalogue}
\label{appendix:greco}

A similar catalogue to ours was produced by \cite{Greco2018}, who used the first data release (DR\RNum{1}) of the HSC-SSP to search for LSB galaxies over a similar footprint to this study. While the footprint in DR\RNum{1} is slightly smaller than what we have used here, a fairer comparison can be made focusing on a smaller region of sky probed by both studies, in this case we use the 211.5$<$RA [deg]$<$223.5, -2$<$Dec [deg]$<$1 region. We find 83 sources over this region, whereas \cite{Greco2018} found 88. However, only 43 of these sources are common to both catalogues after matching within 10\arcsec. 

\indent The majority of sources we missed are too small to have met our preselection / selection criteria (their catalogue contains galaxies with $r_{e}$$>$2\arcsec, compared to our criteria of $\rec$$>$3\arcsec). Some are very faint and are either missed by our source extraction pipeline or are removed by the ANN selection. In any case, we expect to miss a certain fraction of small, faint objects and this effect is captured by our recovery efficiency estimate.

\indent There are several large sources present in our catalogue that are not in that of \cite{Greco2018}. The most plausible reason for this is the differences in quality and coverage between DR\RNum{1} and DR\RNum{2} of the HSC-SSP and we do not investigate this further. 

\begin{figure*}
	\includegraphics[width=\linewidth]{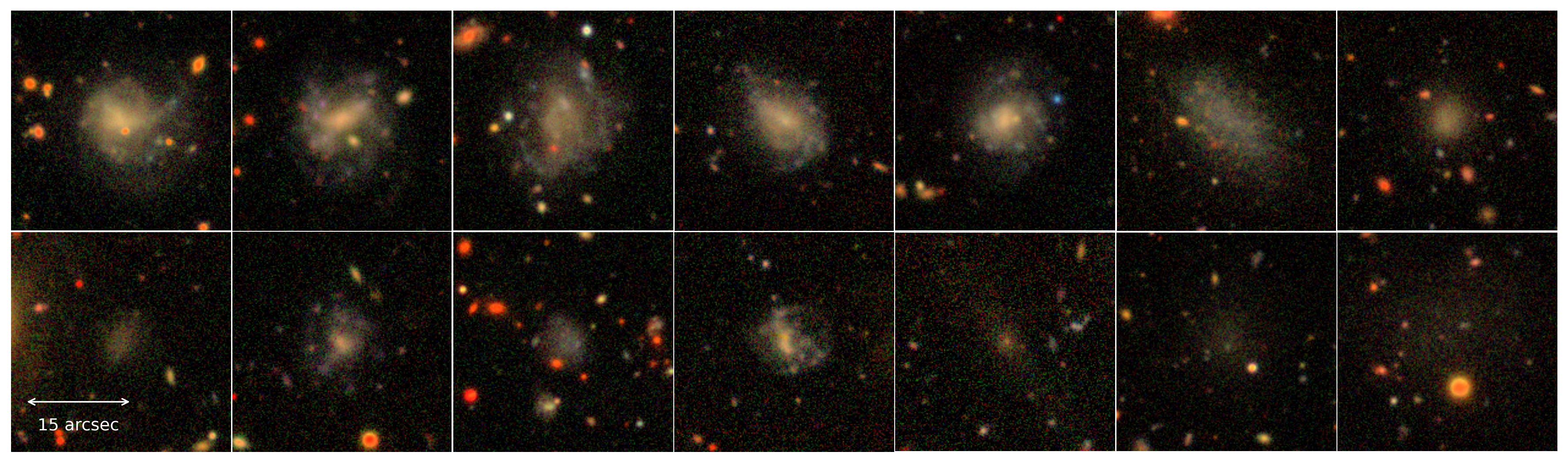}
	\centering
	\caption{Sources in our new catalogue but not that of \protect\cite{Greco2018} (top panels) and  sources in the \protect\cite{Greco2018} catalogue but not ours (bottom panels), after crossmatching the two catalogues in a region common to both surveys. Of the sources missing from our catalogue, many are either too small to pass our preselection or are very faint.}
	\label{figure:sourceplot_greco}
\end{figure*}

\indent Aside from the lower size cut, there are some other important differences between the two catalogues. For example, it is well known that the 128$\times$128 pixel (20\arcsec$\times$20\arcsec) background subtraction in DR\RNum{1} prohibits the accuracy to which LSB structure can be measured. Additionally, \cite{Greco2018} did not use an inclined sky plane while fitting S\'ersic profiles with \Imfit, which we found to be important for recovering unbiased size estimates in this work. They also relied more heavily on human validation of their sources, rejecting 50\% of their sources compared to the 27\% here.

\indent Despite their differences, it is clear that the combination of the two catalogues contains over 1000 LSB galaxies spread over a relatively small footprint. This combined catalogue may prove an invaluable asset to future studies of LSB galaxies.


\end{document}